# Artificial Intelligence in Port Logistics: A Bibliometric Analysis of Technological Integration and Research Dynamics

**Abdelhafid KHAZZAR, *(PhD candidate, MA)***
*Center for Doctoral Studies: Management, Finance, Digitalization and Applies Statistics*
*Faculty of Economics and Management,*
*Abdelmalek Essaadi University, Tétouan, Morocco*

**Yassine SEKAKI, *(Professor, MA)***
*Center for Doctoral Studies: Management, Finance, Digitalization and Applies Statistics*
*Faculty of Economics and Management,*
*Abdelmalek Essaadi University, Tétouan, Morocco*

**Yasser LACHHAB, *(PhD candidate, MA)***
*Center for Doctoral Studies: Management, Finance, Digitalization and Applies Statistics*
*Faculty of Economics and Management,*
*Abdelmalek Essaadi University, Tétouan, Morocco*

**Said EL-MARZOUKI, *(Professor, MA)***
*Center for Doctoral Studies: Management, Finance, Digitalization and Applies Statistics*
*Faculty of Economics and Management,*
*Abdelmalek Essaadi University, Tétouan, Morocco*

| | |
|---|---|
| **Correspondence address :** | Faculty of Legal, Economic and Social Sciences<br>Av. Hassan II, Martil<br>Université Abdelmalek Essaâdi<br>Morocco (Tétouan)<br>93150<br>+2120539687086 |
| **Disclosure Statement :** | The authors declare that they have not received any financial support that could have influenced the objectivity of this study. They take full responsibility for any potential plagiarism, the use of artificial intelligence in the writing process, as well as for the results presented in this article. |
| **Conflict of Interest :** | The authors report no conflicts of interest. |
| **Cite this article :** | KHAZZAR, A., SEKAKI, Y., LACHHAB, Y., & EL-MARZOUKI, S. (2025). Artificial Intelligence in Port Logistics: A Bibliometric Analysis of Technological Integration and Research Dynamics. *International Journal of Accounting, Finance, Auditing, Management and Economics*, *6*(10), 662–679. |
| **License** | **This is an open access article under the CC BY-NC-ND license** |

# Artificial Intelligence in Port Logistics: A Bibliometric Analysis of Technological Integration and Research Dynamics

## Abstract

The paper explores the transformation of port logistics operations with artificial intelligence (AI) during the port transformation into a smart port. The research integrates capabilities-based resource analysis and dynamic capabilities with sociotechnical implementations of technologies and resilience approaches of complex systems under disruptions. The system applies robust data infrastructures (AIS/IoT streams and data quality and interoperability) to propel analytical and AI modules (forecasting and optimization and anomaly detection) that become effective once integrated with sufficient governance systems and trained personnel and operational processes to transform planning and safety and sustainability operations.

It applies Scopus bibliometric research to analyze 123 articles using a systematic approach with both a search protocol and a document screening and duplication verification. It incorporates annual behavior and distribution of author and country performance analysis with science mapping techniques that explore keyword relation and co-citation and bibliographic coupling and conceptual structuring tools that construct thematic maps and multiple correspondence analysis with community detection while applying explicit thresholding and robust tests.

The research connects AI applications to smart port domains through specific data-to-impact pathways while providing a method for bibliometric analysis that enables future updates. The research presents a step-by-step approach for data readiness followed by predictive and optimization implementation and organizational integration. The paper supports public policy through recommendations for data sharing standards and complete environmental benefit assessments. The research proposes a future study plan which combines field-based testing with multiple port assessments to enhance both cause-effect understanding and research applicability.

## 1. Introduction

Artificial intelligence (AI) transforms maritime logistics and port operations through predictive and prescriptive decision-making systems which now operate autonomously for berth and yard planning and gate management and safety and environmental performance. The 80% of global merchandise volume transported by sea requires minimal improvements in turnaround times and resource allocation and emission reduction to create substantial benefits for worldwide supply chains (Notteboom & Rodrigue, 2021). Real-time information and coordinated data-driven control systems become more valuable for ports because they face rising challenges from congestion and scheduling problems and energy efficiency needs and changing regulatory requirements (Parola & Satta, 2020; Lim & Kim, 2023). The data infrastructure of Industry 4.0 through cyber-physical systems and dense IoT sensing and high-fidelity digital twins enables the deployment and governance of AI modules for forecasting and optimization and anomaly detection (Lee et al., 2015; Heilig et al., 2017; Chen et al., 2022; Wu et al., 2023; Zhang et al., 2022). The technological stack enables port operations to achieve end-to-end ETA prediction for congestion prevention and dynamic berth–crane–vehicle allocation and condition-based maintenance and safety analytics through digital-twin platforms that allow counterfactual testing and decision rehearsal before operational deployment. AI systems now support environmental goals through their implementation of eco-routing and just-in-time arrival and multi-objective operational policies that optimize service levels and costs and minimize emissions (Bouman et al., 2017; Psaraftis & Kontovas, 2020; IMO, 2020).

The academic field shows rapid development yet scholars have not unified their research into a cohesive body of knowledge. The current body of research investigates blockchain and IoT and digital twins and algorithmic methods independently yet it does not study the complete transformation of port logistics systems by AI and the organizational elements that influence successful large-scale deployment (Chang & Chen, 2021; Luo & Shin, 2022; Zhu & Wang, 2024). The research gap exists in our target corpus and recent sustainability-focused port studies because they show the need for complete system assessments that move beyond evaluating individual tools to assess digital readiness and governance and workforce adaptation. The successful execution of digital transformation requires organizations to implement technology systems which match their operational abilities and security measures and their particular training needs. The research performs a field-based AI port logistics study through bibliometric analysis which uses open methods for achieving reproducibility. The research team initiated their systematic process by extracting 123 peer-reviewed documents from Scopus followed by duplicate removal and document verification and visualization analysis. The research design combines statistical indicators with science-mapping approaches which include keyword co-occurrence and co-citation and bibliographic coupling and conceptual-structure mapping (MCA) through Bibliometrix software using Biblioshiny. The research method produces quantitative information and visual representations which demonstrate the field's publication output and its knowledge base and thematic progression. The research data shows three main patterns in the study results. The number of research papers about AI and digital-twin applications in maritime operations has shown rapid growth since 2020 because decision-support and simulation systems gained importance for ports and AI penetration increased throughout maritime industries; our study examines publications from 1996 to 2024 with a significant increase in recent years. The research core themes include simulation/optimization and decision support but sustainability topics such as eco-routing and energy-aware dispatching and just-in-time arrival have become vital components which integrate with digital twins as decision platforms. The current barriers to deploying AI at port scale stem from ongoing issues with data quality and semantic interoperability and data/model asset governance and model transferability between ports. The observed patterns support the theory-based interpretation which follows in the subsequent sections.
The research investigates AI-driven port transformation through three conceptual stages which align with the findings from our literature review. Strategic level analysis through capability-based views shows how data assets and domain-specific models and MLOps infrastructure and human capital evolve into enduring competitive advantages through strategic reorganization under uncertain conditions. The combination of socio-technical alignment with routinization processes determines why terminals using identical tools achieve varying success because digital success requires technology-process alignment and skill development in analytics and cybersecurity and sustainability management. The adoption frameworks (TOE/DOI) and institutional pressures (regulatory and normative) determine the speed and manner of implementation while interoperability standards and data-sharing governance mechanisms play a central role in the process. The implementation of strategic diagnostics (SWOT–CAME) and competency development prove essential for port ecosystems to achieve successful technology deployment according to recent sustainability research.The research aims to achieve three main objectives: (i) Develop a reproducible AI in port logistics science map which displays production patterns and origins and distribution areas and conceptual frameworks (ii) The research combines technological development paths with business operational potential and system limitations through a structured theoretical framework (iii) The research develops practical and policy recommendations which include data-sharing protocols and competency training and complete environmental impact assessments to ensure AI systems meet decarbonization targets. The research achieves its objectives through a documented Bibliometrix/Biblioshiny workflow which combines quantitative structure analysis with qualitative interpretation to support ongoing research and future updates.
The article continues with Section 2 which establishes the theoretical framework needed to understand AI functions within port ecosystems. The research methodology section in Section 3 explains how the authors built their corpus and performed science mapping procedures. The research section 4 presents results about production patterns and origin points and participating nations and conceptual frameworks. The final section presents research implications and limitations and future research directions which focus on data readiness and interoperability and governance systems to convert AI pilots into sustainable operational value.

## 2. Theoretical backgrounds

This section outlines the theoretical lenses that explain how artificial intelligence connects with maritime logistics and port operations. Four foundations underpin this integration: digital transformation within port systems, risk management and resilience theory, environmental optimization, and supply-chain visibility.

### 2.1. Digital Transformation and the Smart Port Paradigm

The digital transformation of seaports constitutes a fundamental aspect of the overall smart logistics approach. AI together with big data analytics and IoT powers Smart Ports to automate and enhance port operations (Heilig, Lalla-Ruiz, & Voß, 2017). AI-based predictive maintenance techniques have decreased machine breakdowns in unexpected operations (Liu, Zhang, & Li, 2021) while smart scheduling systems cut down port berthing times and turnaround durations (Garcia & Gonzalez, 2020).
The implementation of digital twins allows monitoring physical port assets through virtual replicas that enable real-time data tracking and anomaly detection alongside scenario simulation capabilities (Tao et al., 2019). Wu et al. (2023) show that maritime logistics digital twins increase situational understanding and establish an operational framework for making proactive decisions in changing environments. The innovative approaches unite as part of the Logistics 5.0 framework which focuses on human-centered automation together with sustainable intelligence.

### 2.2. Risk Management and Resilience Theory

AI strengthens resilience by supporting real-time risk assessment, disruption prediction, and decision-making. Maritime supply chains can suffer cascading effects from demand swings, weather shifts, and geopolitical tensions, which push risks through tightly linked networks (Ivanov, 2020). In this setting, AI's forecasting, detection, and optimization capabilities help ports anticipate shocks and coordinate more effective responses. In practical terms, early-warning tools for port congestion can smooth truck arrivals and improve gate operations (Zhang & Lee, 2022). The research conducted by Lütjen et al. (2021) demonstrates that machine learning systems require historical data to train models which enhance port operational resilience through container misplacement and customs bottleneck predictions.

### 2.3. Sustainability and Environmental Optimization

AI tools help the maritime industry achieve environmental standards through operational tools which support goal alignment. The system generates speed adjustments and route plans through emission data analysis and weather forecasts and ship performance modeling to achieve reduced fuel consumption and lower carbon emissions (Psaraftis & Zis, 2018).
The combination of AI technologies helps users comply with Emission Control Area regulations through the detection of irregularities in emissions while suggesting optimal navigation routes for fuel conservation (Bouman et al., 2017). The models use optimization algorithms which include genetic algorithms and neural networks (Shukla & Jharkharia, 2013).

### 2.4. Theoretical Foundations of Sustainable and Digital Port Logistics

Three separate theoretical frameworks support the development of port digitalization and sustainability through their combined analysis. The SWOT–CAME approach serves as a strategic management framework which enables organizations to evaluate their digital readiness and create specific business strategies (Correct, Adapt, Maintain, Explore). The research on Spanish ports uses SWOT–CAME analysis with Delphi expert validation to determine the most important implementation paths (cybersecurity and workforce development and public-private partnerships) that connect new technologies to sustainability targets (González-Cancelas et al., 2025).
The strategic decision-making framework for eco-efficient port operations integrates the Metaverse and digital twins with AI and IoT as essential components.
The human-capital and competency approach focuses on the organizational abilities which support digital-sustainability strategy implementation. A mixed-methods study developed a competency framework for maritime and port logistics which places strategic supply-chain optimization and data analysis and port operational technical skills and digital transformation capabilities and sustainable project management and renewable energy knowledge at the forefront (Lopes et al., 2025).
The research evidence supports capability-based competitive advantage theory because organizations and ports build and transform their digital and green competencies to handle environmental and regulatory changes which results in continuous port ecosystem performance enhancement. (Synthesis based on the competency evidence above.)
The operations and socio-technical approach unites data-focused approaches with system-level design to achieve both operational efficiency and environmental sustainability. The combination of machine learning with discrete-event simulation produces predictions that match actual results with high precision which optimizes equipment waiting times and empty moves to decrease port duration and lower carbon emissions (Benghalia et al., 2025).

The data-driven decision-making approach supports the strategic and human-capital frameworks to create a comprehensive theoretical framework which unites strategic diagnosis with capability development and model-based optimization to enhance sustainable digital port logistics.

## 3. Research Design

A bibliometric analysis framework based on reproducible systematic procedures helped us investigate AI applications in maritime logistics and port systems through thematic development and scholarly growth. The research design consists of four separate methodological components which are described in this section.

### 3.1. Data Source and Search Strategy

The research used Scopus as its data source because it provides extensive peer-reviewed content in engineering and logistics and computer science and maritime studies which are essential for AI port logistics research and its standardized metadata standards allow for reproducible bibliometric analysis. The title–abstract–keyword search on April 9 2025 used "artificial intelligence" and "port" and "logistics" to retrieve 146 records. The search included all time periods to study the complete evolution of the field instead of focusing on recent developments. The search results included 146 records.

***Figure 1: Diagram of the Selection Process***

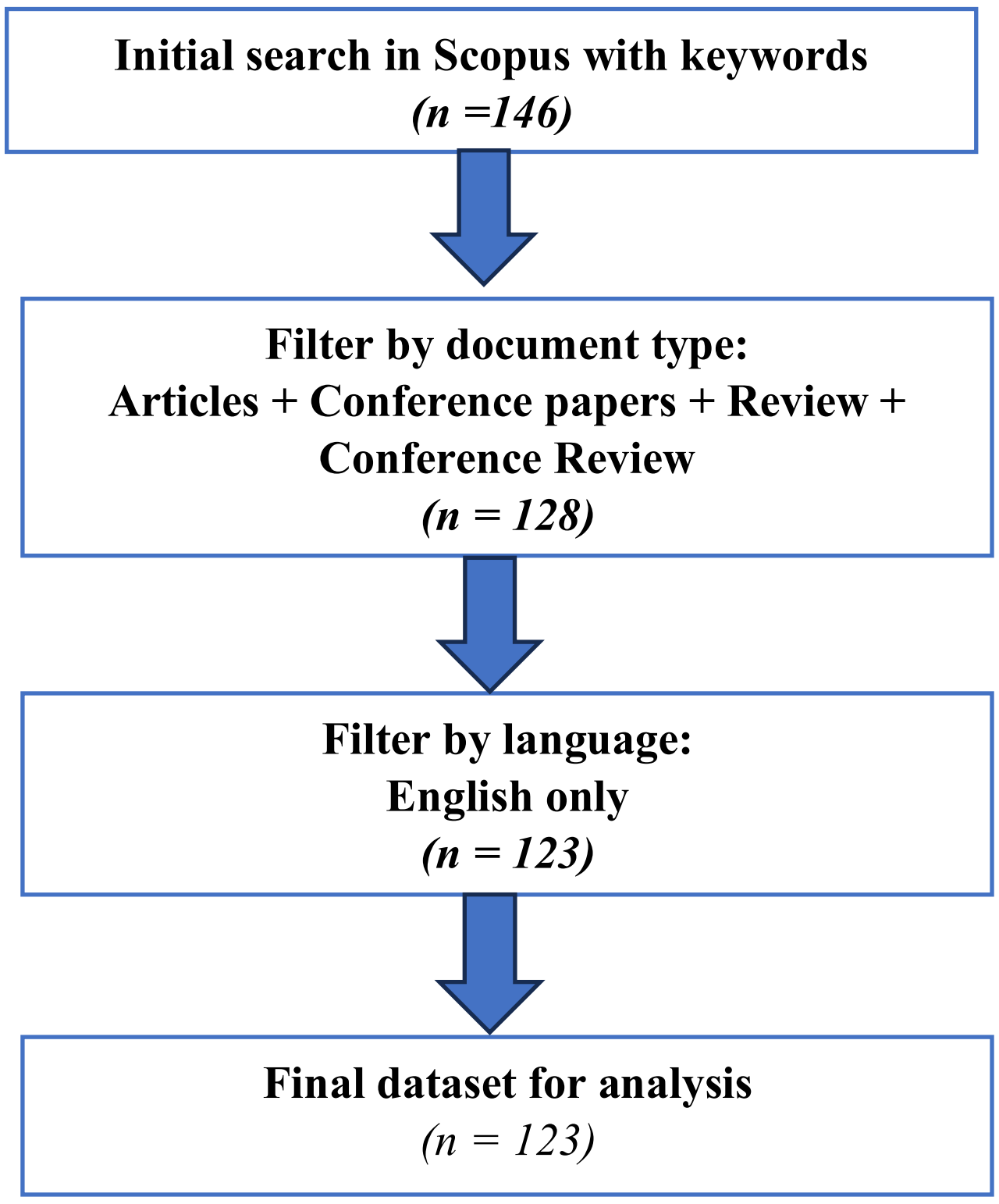

***Source: Created by authors***

The selection process for the corpus followed a two-stage method which met both methodological requirements and analytical needs for bibliometric standards. The dataset contained only journal articles and conference papers and reviews and conference reviews which made up a total of 128 items. The document-type filter was used because these formats undergo peer-review and contain complete structured metadata which science-mapping techniques require but paratexts like editorials and notes do not provide sufficient scholarly value. The analysis focused on English-language documents (n = 123) because this choice ensured text preprocessing consistency and better interpretability while English functions as the primary scholarly communication language in this field.
The study included records that showed clear connections between artificial intelligence and port operations and container terminal logistics and provided an informative abstract. The analysis excluded duplicate records and non-English items and editorials and notes and records without abstracts and studies where AI-port-logistics connections played a minor role. The complete process from initial retrieval to the final analytical corpus of 123 studies appears in Figure 1 (Diagram of the Selection Process).

### 3.2. Data Processing and Cleaning

Bibliographic metadata—including title, abstract, keywords, authors, sources, citations, and affiliations—were exported from Scopus in BibTeX format. The data was then imported into the Bibliometrix R-package for preprocessing. Data cleaning involved standardizing author names, merging keyword variants, and removing irrelevant entries (e.g., documents without abstracts or with inconsistent formatting).
This step also ensured compatibility with the Biblioshiny interface, enabling interactive analysis and visualization. Outliers and noise (such as misclassified document types or non-English texts) were manually reviewed and excluded.

### 3.3. Analytical Framework and Tools

Our analysis combined descriptive bibliometric indicators with advanced network mapping techniques. Key metrics included publication trends, most prolific authors, top contributing countries, citation patterns, and keyword co-occurrences.
We used co-word analysis to uncover conceptual structures and co-citation analysis to identify the most influential publications. Thematic evolution analysis helped track how core topics shifted over time. This methodological triangulation ensures robustness and comprehensiveness in capturing the intellectual structure of the field.

### 3.4. Visualization and Interpretation

Biblioshiny's visualization tools were leveraged to generate strategic diagrams, thematic maps, word clouds, and Sankey plots. These visual aids facilitated pattern recognition and supported our interpretation of the data.
The triangulated findings from Bibliometrix allowed us to identify emerging hotspots (e.g., digital twins, green logistics, AI-based routing), understand inter-author collaboration networks, and assess the maturity of various sub-domains within the research field.
The combination of a systematic review protocol with rigorous data analytics ensures that our study meets academic standards for transparency, reproducibility, and analytical depth.

## 4. Findings and Results

This section presents the empirical results of our bibliometric analysis, derived from the final dataset of 123 publications. The findings are organized into key dimensions to illustrate the structural, intellectual, and conceptual evolution of AI in port logistics.

### 4.1 Annual Scientific Production

The temporal progression of academic interest is captured in Figure 1, which illustrates the annual scientific production related to artificial intelligence, logistics, and port systems from 1996 to 2024. The data show a relatively low and unstable publication rate until 2010. However, a modest upward trend begins around 2011, followed by noticeable fluctuations in the following decade. A significant surge is observed starting in 2021, reaching a peak in 2023 with more than 17 publications. This recent growth reflects the increasing academic interest in the integration of AI technologies within port and logistics environments. The sharp drop observed in 2024 may be due to incomplete indexing for the current year or delayed publication records.

***Figure 1: Annual Scientific Production***

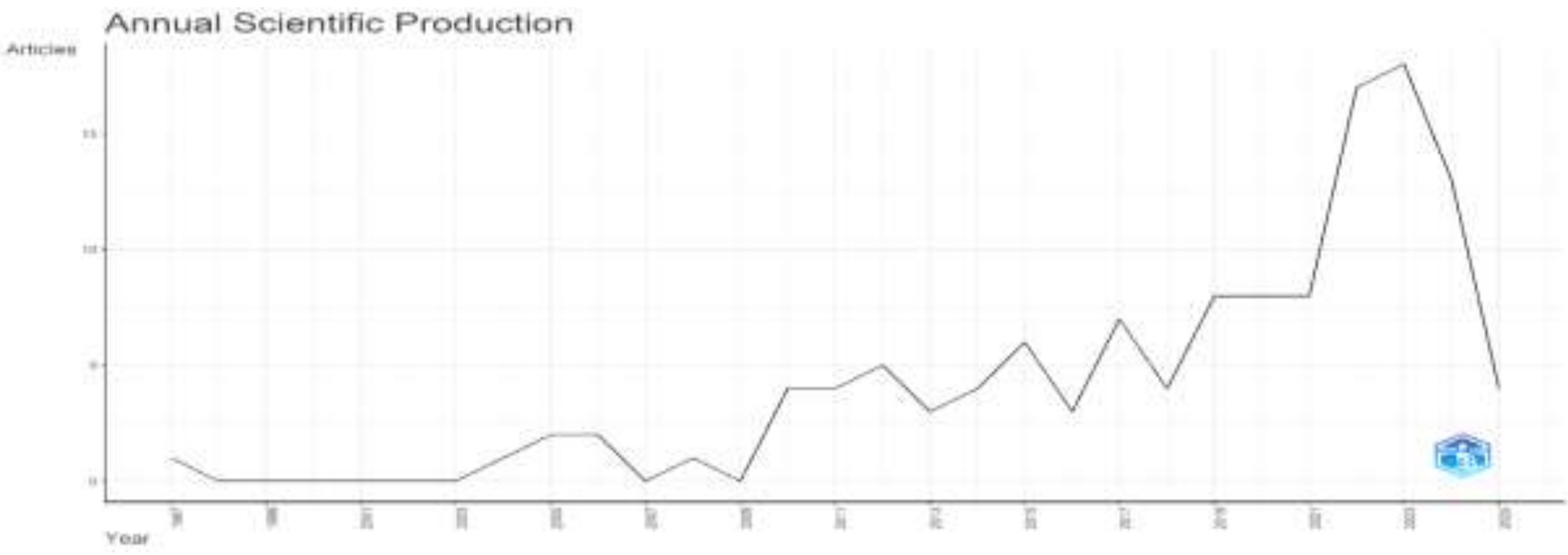

***Source: Created by authors***

## 4.2 Most Relevant Sources

To identify key publication venues, Figure 2 highlights the most relevant sources contributing to the academic literature on artificial intelligence, logistics, and port systems. The Lecture Notes in Computer Science series stands out as the leading source, with 7 published documents, followed closely by Lecture Notes in Networks and Systems (6 documents). These two sources reflect the high technical and computational focus of research in this area. Other relevant venues include Ocean Engineering and the 2011 2nd International Conference on Artificial Intelligence, both contributing 3 papers, indicating a growing interdisciplinary approach that merges maritime logistics with cutting-edge AI innovations. Additionally, journals such as Applied Sciences (Switzerland) and Engineering Proceedings each published 2 documents, underscoring the interest in applied engineering solutions within port logistics contexts. This diversity of sources reflects a broad academic interest cutting across engineering, computer science, and maritime studies.

***Figure 2: Most Relevant Sources***

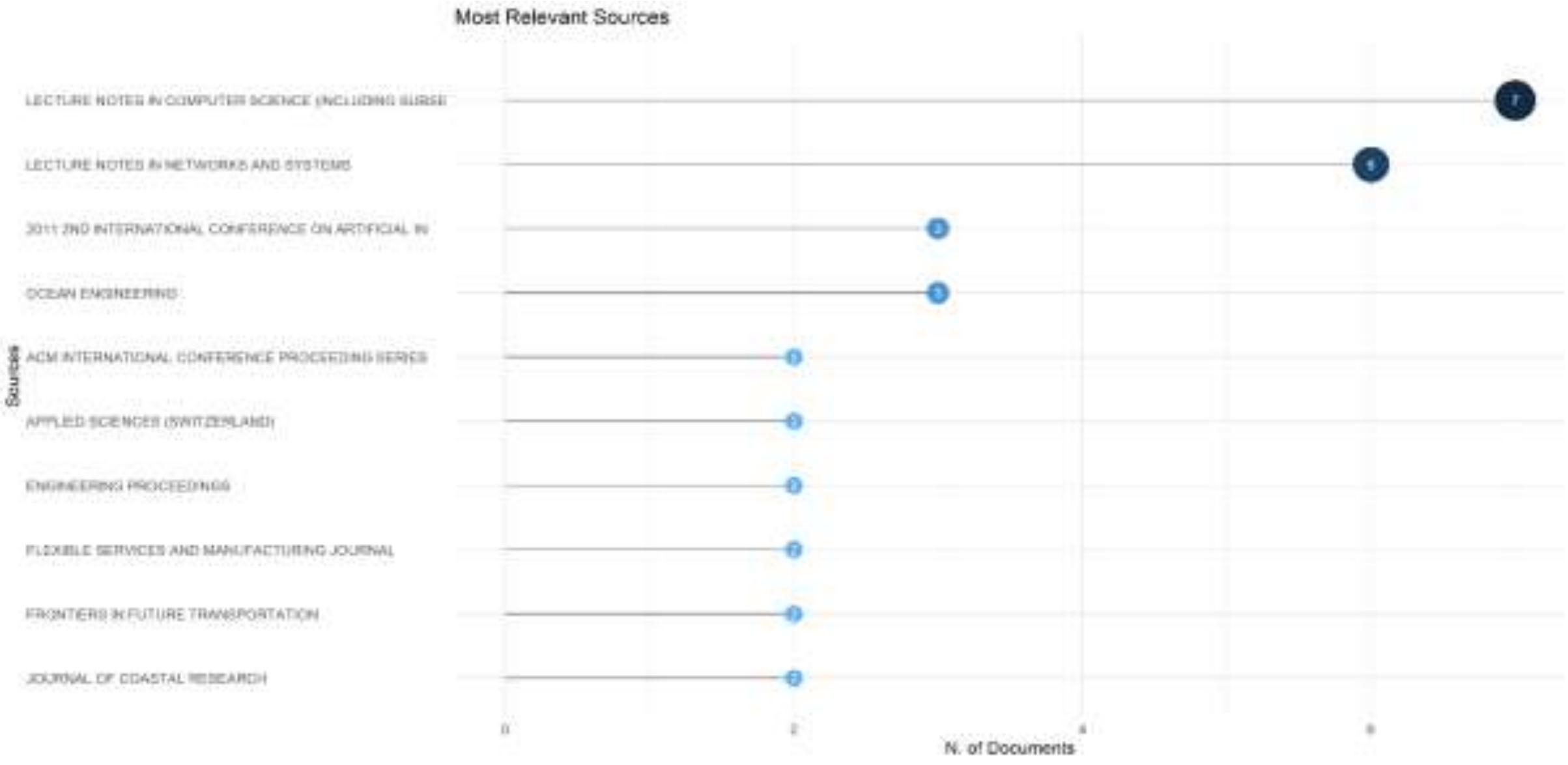

***Source: Authors***

### 4.3 Sources' Production Over Time

As shown in Figure 3, the evolution of source productivity reveals the temporal dynamics of academic interest in artificial intelligence applications in port logistics. Lecture Notes in Computer Science demonstrates a strong and continuous increase in cumulative publications since 2020, indicating a consistent research output in this domain. Similarly, Lecture Notes in Networks and Systems and Ocean Engineering exhibit notable growth from 2021 onwards, reflecting a convergence between AI and maritime engineering topics. Other sources such as Applied Sciences (Switzerland), Engineering Proceedings, and Flexible Services and Manufacturing Journal began contributing to the field more recently, primarily from 2018 to 2023. This distribution confirms that while some outlets have been steady contributors, others have only recently begun engaging with this emerging interdisciplinary theme, likely following the acceleration of digital transformation in port environments.

***Figure 3: Source's Production Over time***

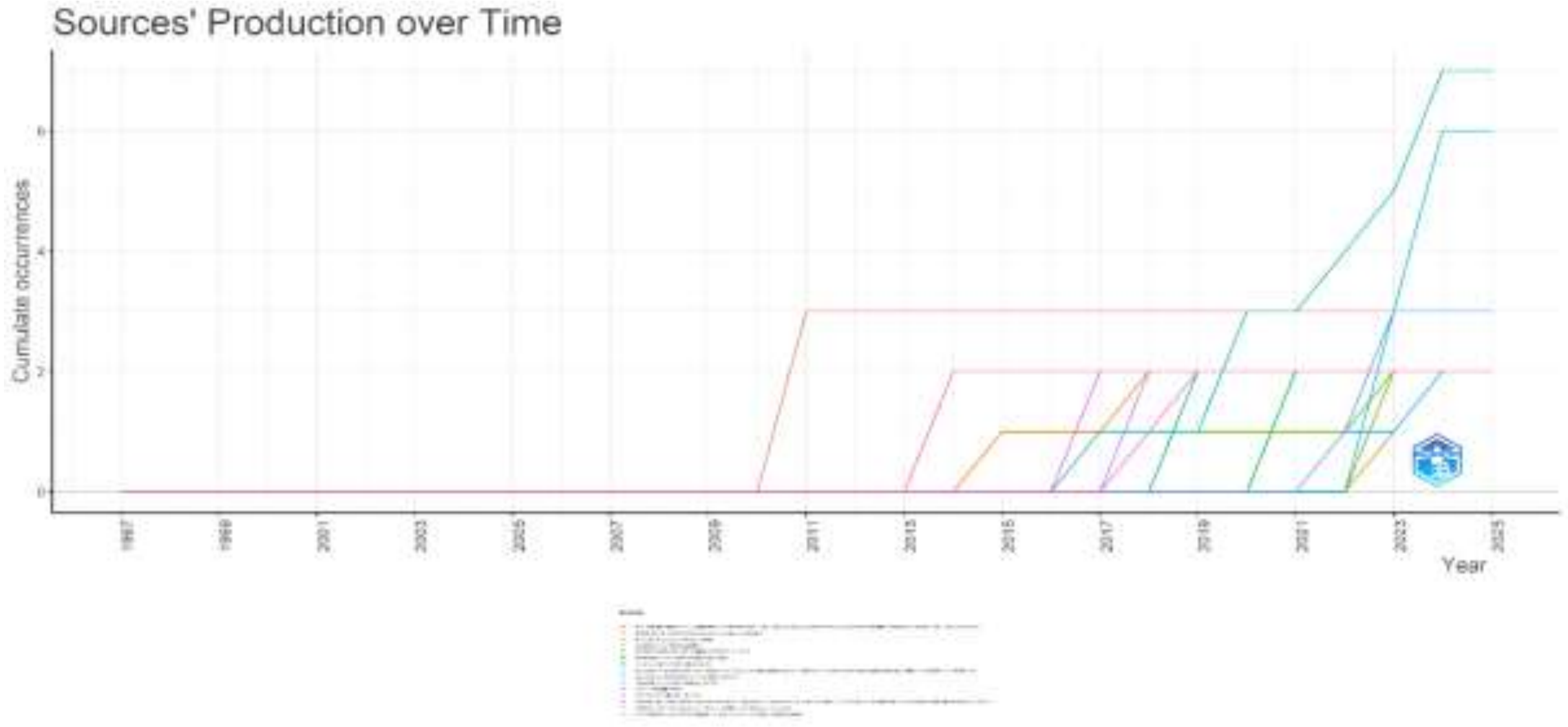

***Source: Authors***

### 4.4 Authors Production Over Time

A focus on individual contributions is offered in Figure 4, presenting the publication activity of the most prolific authors over time in the field of artificial intelligence applied to logistics and port systems. Among the leading contributors, Briano E. and Na Na exhibit long-term involvement in the field, with publications spanning nearly two decades, suggesting sustained research activity. Authors such as Iacobellis G., Lalla-Ruiz E., and Fanti M.P. show more recent but impactful engagement, with multiple contributions concentrated between 2013 and 2023. The size of the bubbles indicates the number of articles, while the color intensity represents the total citations per year (TC/year), highlighting that Iacobellis G. and Fanti M.P. have not only produced multiple papers but have also achieved higher citation rates. This distribution suggests a mixture of veteran researchers with long-term interest in the topic and newer authors contributing to emerging subfields such as smart port operations, AI optimization algorithms, and intelligent logistics systems.

***Figure 4: Authors Production Over time***

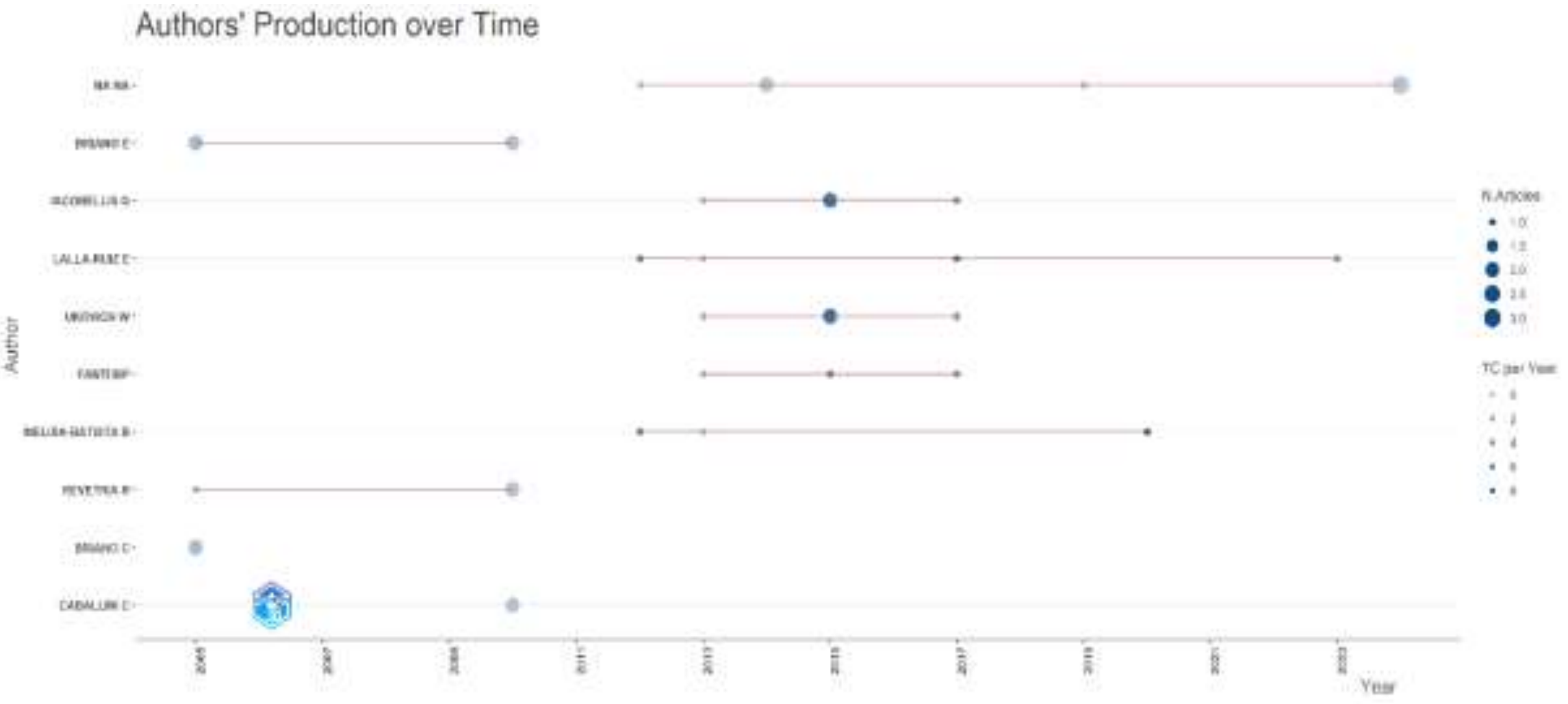

***Source: Authors***

## 4.5 Affiliations' Production Over Time

Institutional engagement in the field is depicted in *Figure 5*, which displays the research output by institutional affiliation. The University of Szczecin stands out as the most consistent contributor, with a steady increase in publications from 2012 to 2023, reaching a total of 5 documents. This highlights its leading role in the academic development of AI and smart port research. Other institutions such as the University of Catania, Zhejiang University, and the University of Trieste have demonstrated increasing interest in the subject, with publications beginning to appear more prominently from 2020 onward. The Maritime University of Szczecin also shows recent engagement in this area. The recent emergence of multiple affiliations since 2020 reflects a growing institutional commitment to digital innovation in port and logistics systems, likely in response to global trade disruptions and the digital transformation of supply chains.

***Figure 5: Affiliations Production Over time***

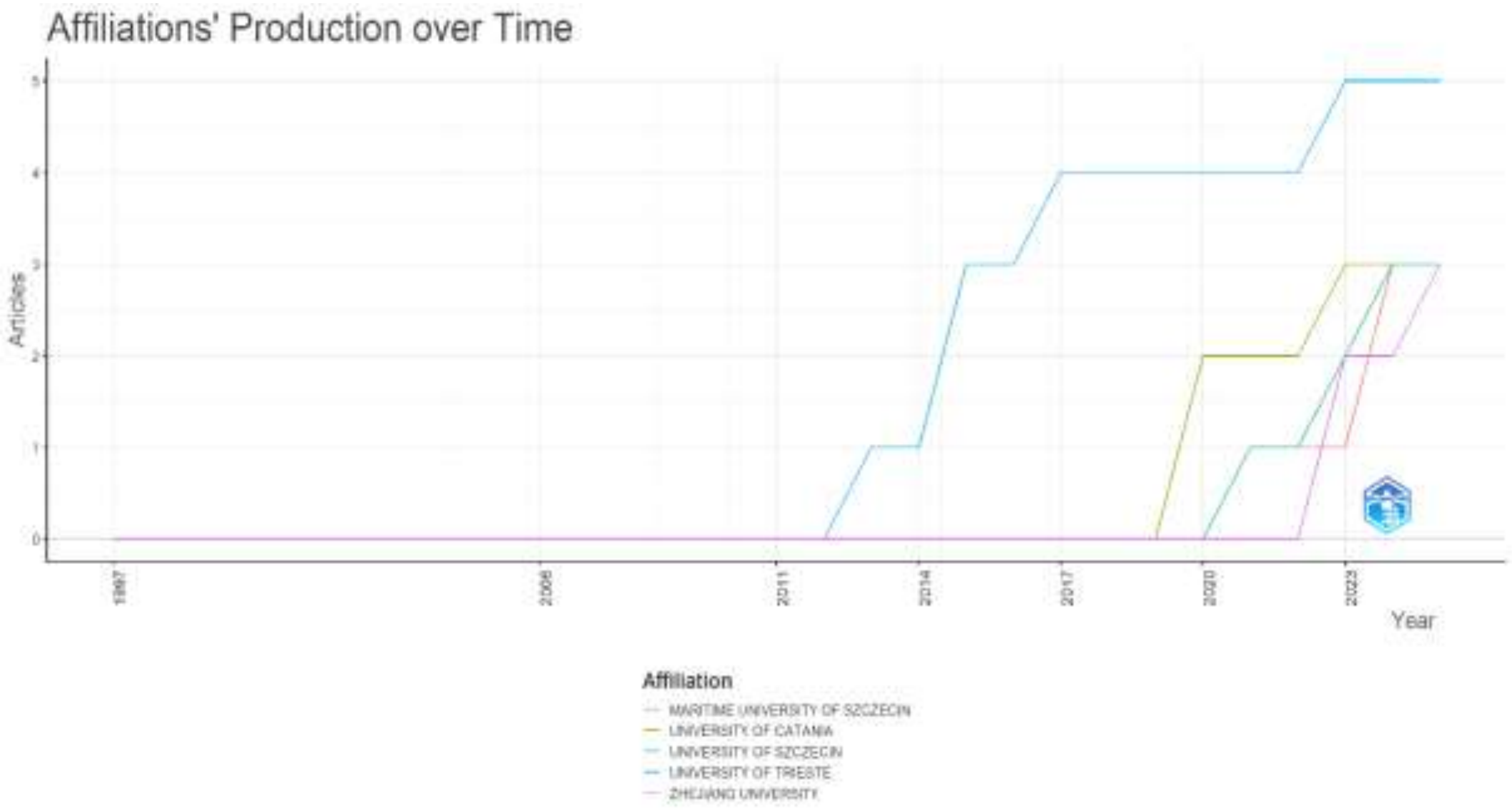

***Source: Authors***

### 4.6 Corresponding Author's Countries

Geographic patterns of corresponding authorship are examined in Figure 6, illustrating the distribution of corresponding authors by country, highlighting both national (SCP – Single Country Publications) and international (MCP – Multiple Country Publications) collaborations. China leads significantly in terms of corresponding authorship, with a notable dominance of single-country publications, indicating a strong domestic research effort in the field. Italy and Spain follow, also with a predominance of national contributions. Korea and Australia show balanced output, while Belgium, Morocco, and Portugal are emerging contributors, suggesting growing academic engagement in smart logistics and port digitalization. Notably, the presence of countries like Germany, the USA, and the UK with multiple-country collaborations reflects their role in international research networks. Overall, the data highlight both the geographic concentration of research in specific regions and the increasing global interest in AI-driven port logistics.

*Figure 6: Corresponding Author's Countries*

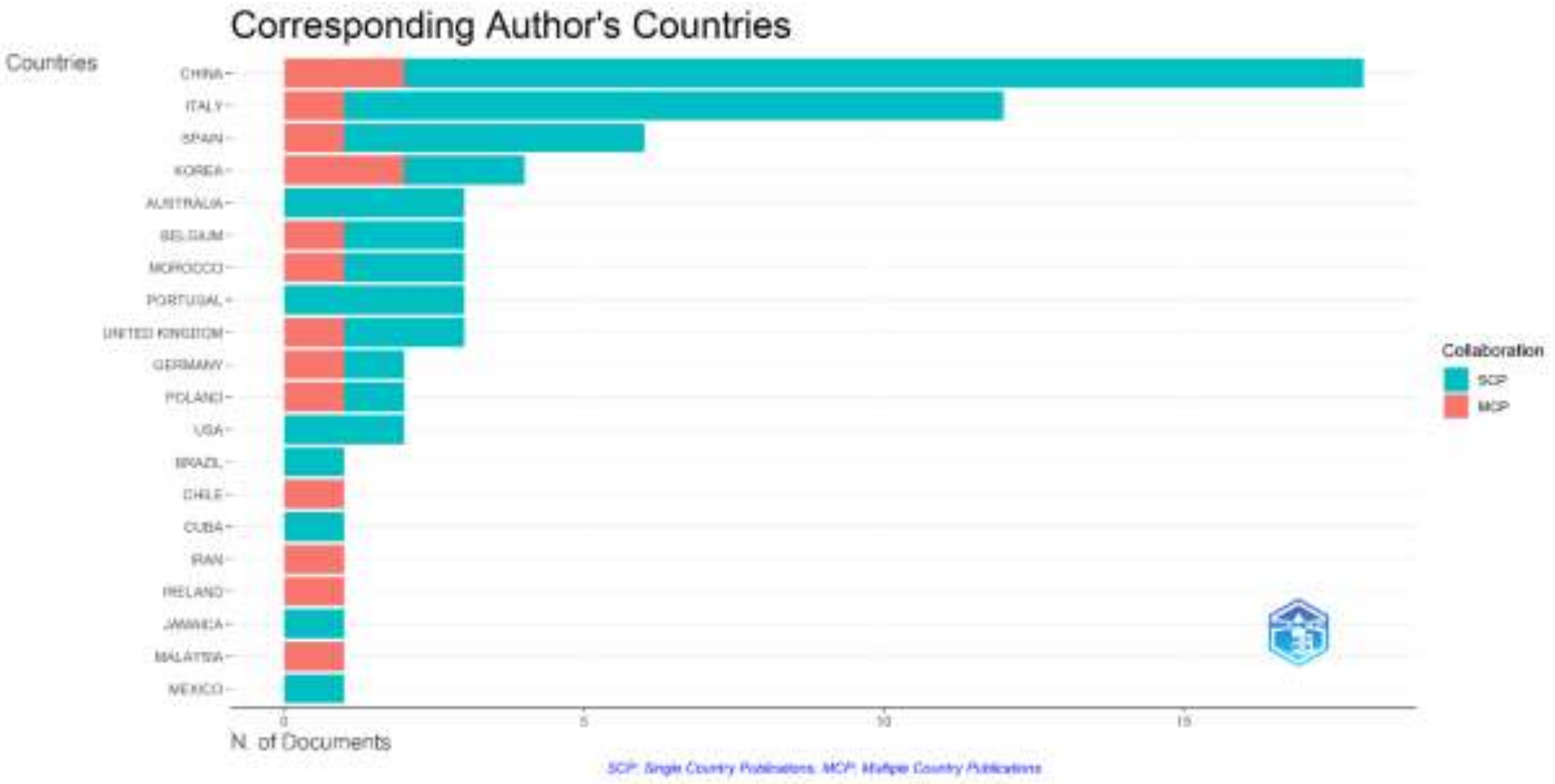

***Source: Authors***

### 4.7 Scientific Production by Country

The global distribution of scholarly output is presented in Figure 7, showcasing the countries with the most contributions. The map highlights that China leads the field with the highest number of publications, reflecting its strategic investment in AI and smart port infrastructure. Other major contributors include India, Australia, Italy, Spain, and the United States, all shaded in darker tones, indicating substantial research output. European countries such as Germany, France, and Portugal, as well as South Korea and Japan, also demonstrate strong engagement in the field. Notably, emerging economies including Brazil, Morocco, and Indonesia are represented, pointing to a growing global interest in applying AI technologies to enhance port operations and logistics performance. This wide distribution suggests that digital transformation in port systems is a truly international concern, aligning with global trade and sustainability agendas.

***Figure 7: Scientific Production by Country***

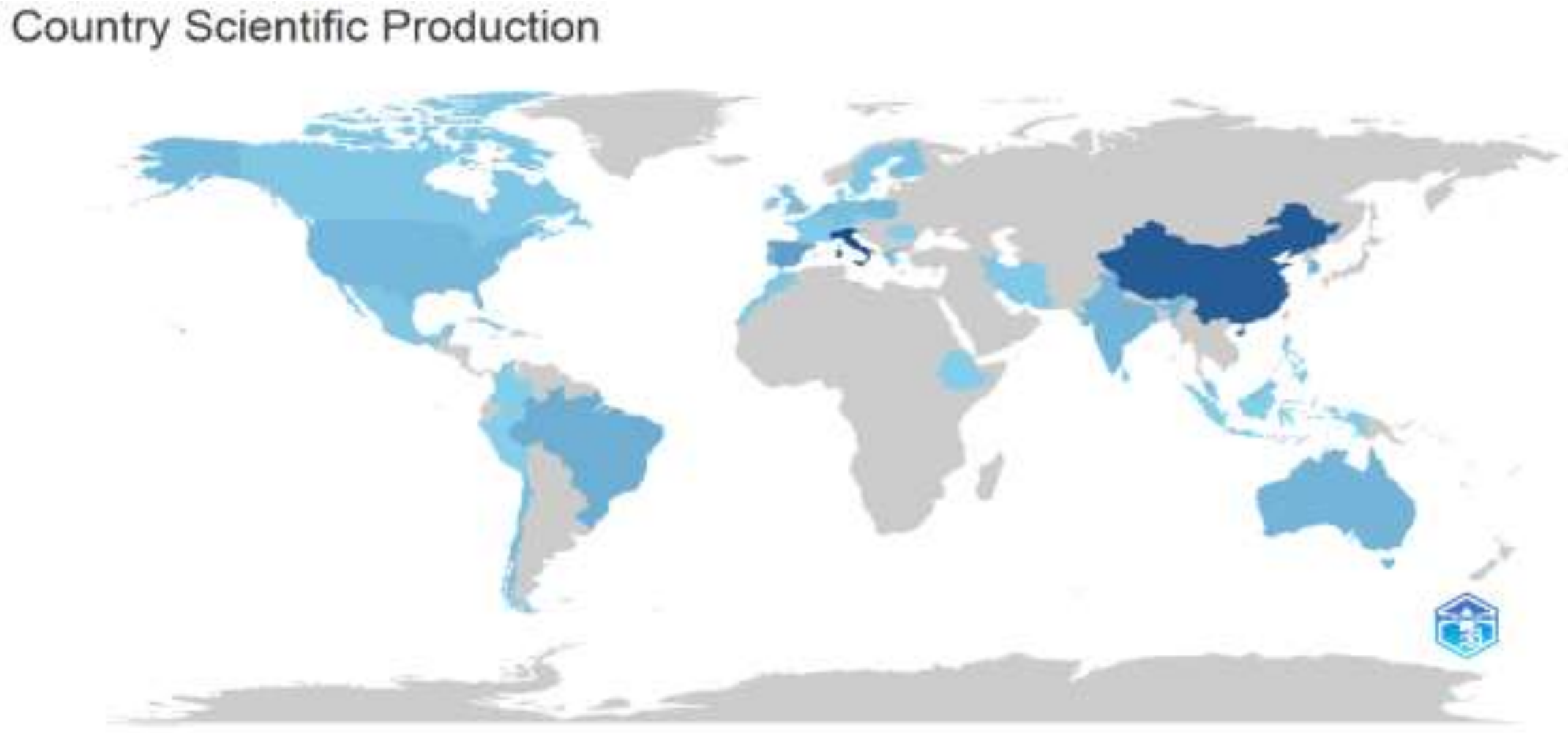

***Source: Authors***

## 4.8 Most Cited Countries

Citation impact by country is summarized in Figure 8, identifying the countries whose publications have received the highest number of citations in the field of artificial intelligence applied to logistics and port systems. Italy ranks first with 257 citations, followed by Spain (180) and the United Kingdom (134), indicating that research originating from these countries has had substantial academic influence. China, despite being the most productive country in terms of publication volume, is fourth in citations (91), suggesting a slightly lower citation impact per article. The United States and Germany follow, with 87 and 58 citations respectively. Countries like Chile, Korea, Australia, and Mexico also appear in the ranking, showing that impactful research is emerging from diverse regions. These figures highlight not only leadership in research output but also visibility and recognition within the academic community.

***Figure 8: Most Cited Countries***

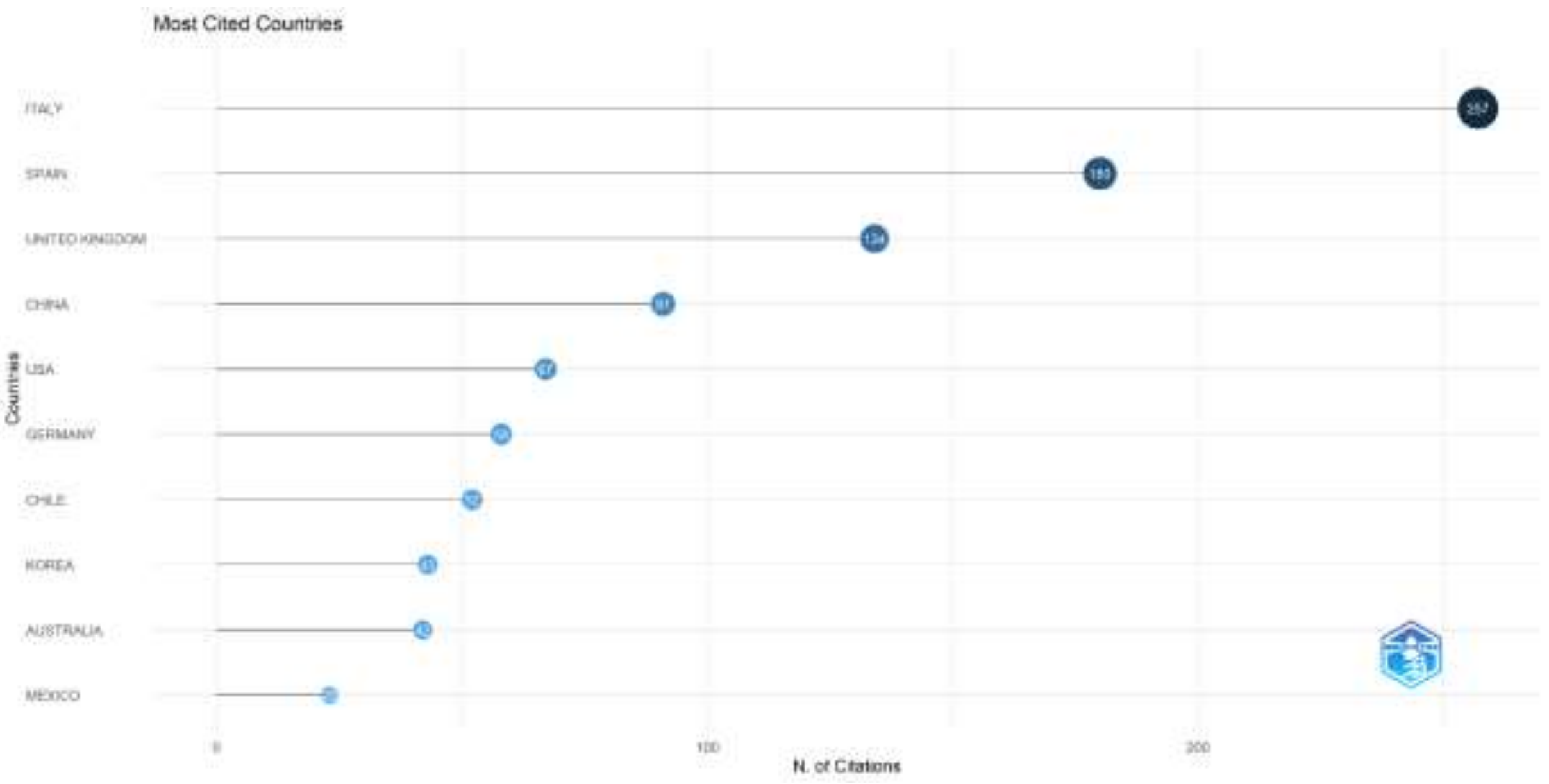

***Source: Authors***

## 4.9 Words' Frequency Over Time

Trends in thematic focus are revealed in Figure 9, which illustrates the frequency of key terms over time, highlighting the evolution of thematic interest in AI and port logistics research. The term Artificial Intelligence demonstrates the highest and most enduring growth since 2010 because AI has become essential for optimizing port operations and supply chain management systems. The research maintains its focus on port environment efficiency and planning and intelligent automation through the repeated appearance of "Container Terminal" and "Decision Support Systems" and "Logistics" terms. The appearance of "Ports and Harbors" and "Railroad Yards and Terminals" and "Ships" in the data indicates rising interest in combining different transportation modes. The rising frequency of keywords during the last ten years indicates that AI-based logistics management has become a vital strategic element for the worldwide maritime industry.

***Figure 9: Words' Frequency Over Time***

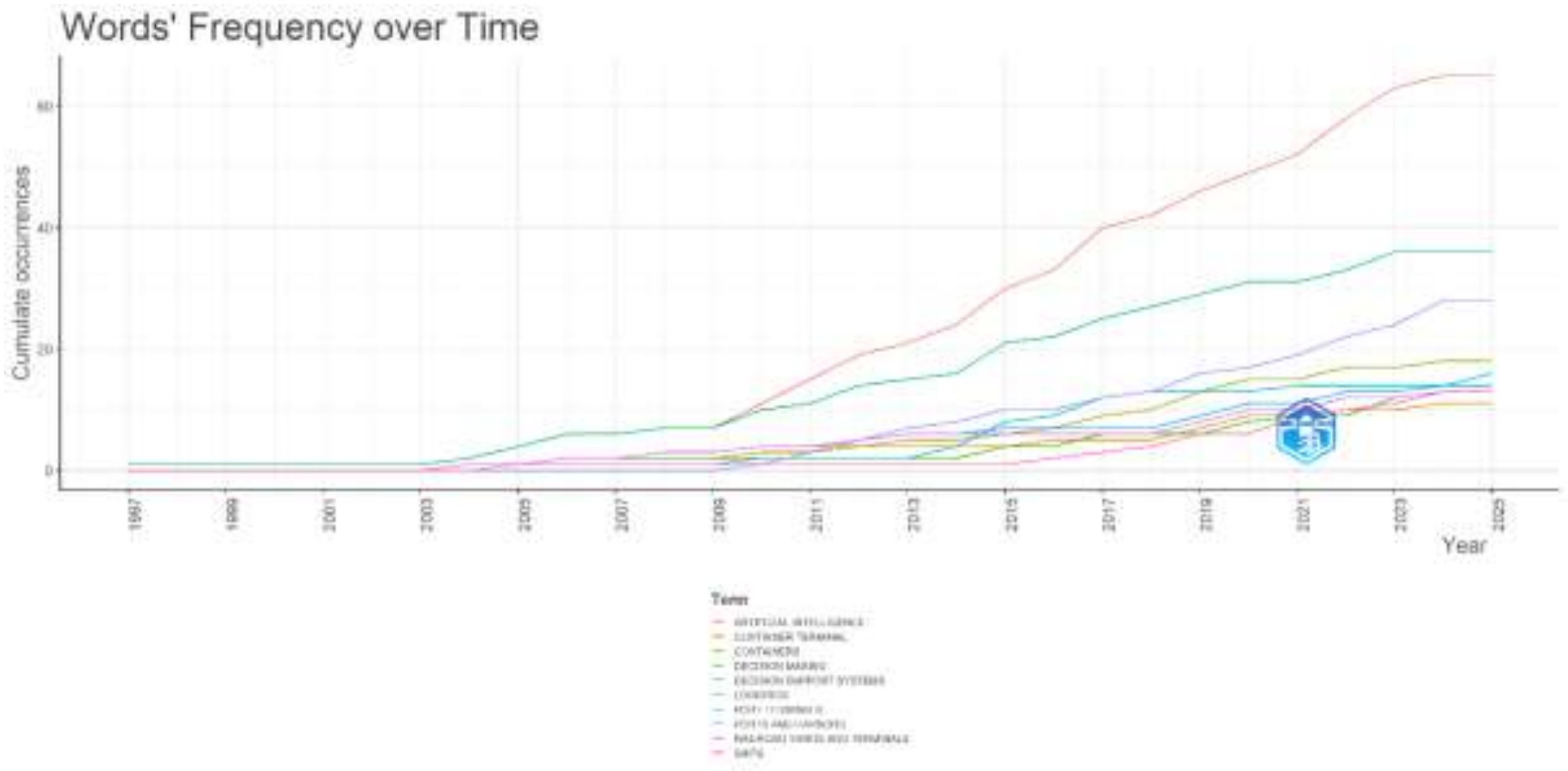

***Source: Authors***

## 4.10 Word Cloud

The most common terms in artificial intelligence and port systems and logistics research appear in Figure 10 as a visual summary. The research field places artificial intelligence at its core because it stands as the dominant keyword according to the visual representation. The research field focuses on decision support systems and ports and harbors and logistics and containers and container terminals because these elements represent the core operational and technological aspects of port logistics that AI enhances. The research focuses on intermodal and global logistics through its examination of railroad yards and terminals and freight transportation and supply chains. The research demonstrates its digital innovation focus through its use of optimization and simulation and machine learning terminology. The visualization demonstrates how AI connects different parts of port logistics systems while showing the diverse range of current research activities.

***Figure 10: Word Cloud***

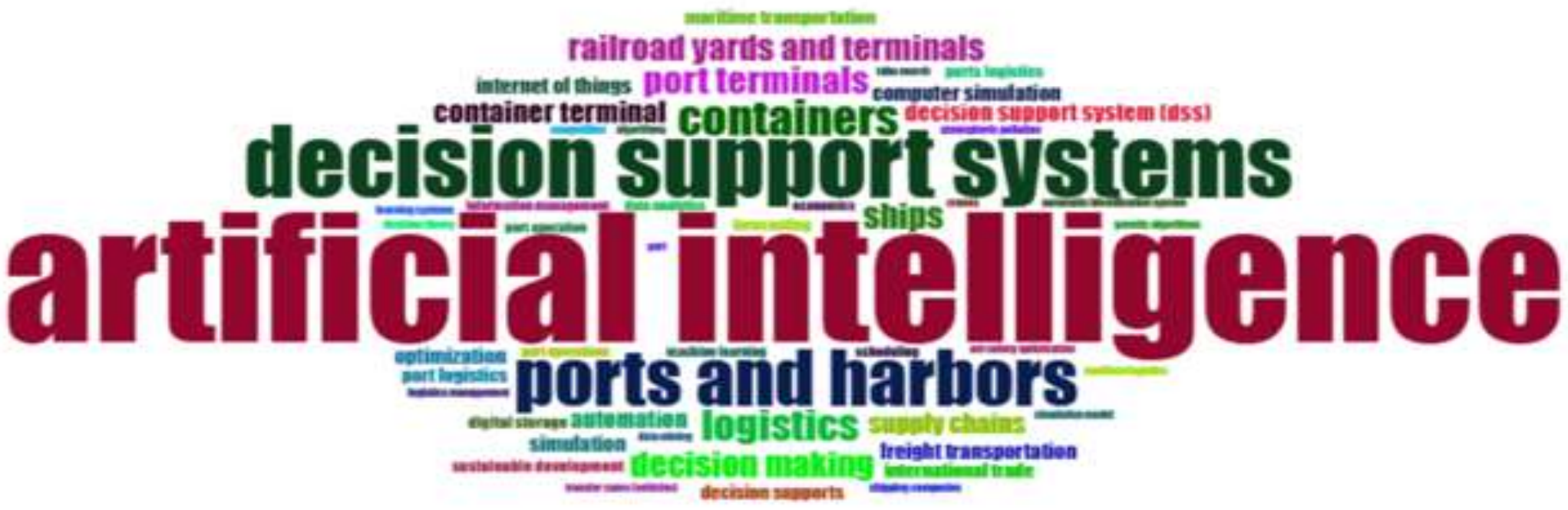

***Source: Authors***

## 4.11 Trending Topics

Emerging areas of interest are captured in Figure 11, highlighting trending topics based on keyword evolution. Among the most recent and frequently discussed terms are "maritime transportation", "port operations", "sustainable development", and "ports logistics", which have gained prominence especially from 2020 onwards, reflecting the increased attention to smart and sustainable port ecosystems. Terms such as "decision support systems", "optimization", "supply chains", and "freight transportation" have shown consistent interest over the past decade, highlighting a strong focus on performance enhancement through AI-enabled solutions. Notably, foundational terms like "computer simulation", "simulation", and "artificial intelligence" began trending earlier and remain central, indicating the persistent importance of modelling and intelligent technologies. This trend landscape reveals both the evolution of core technical topics and the emergence of socio-environmental concerns such as sustainability and global trade integration within port logistics research.

***Figure 11: Trending Topics***

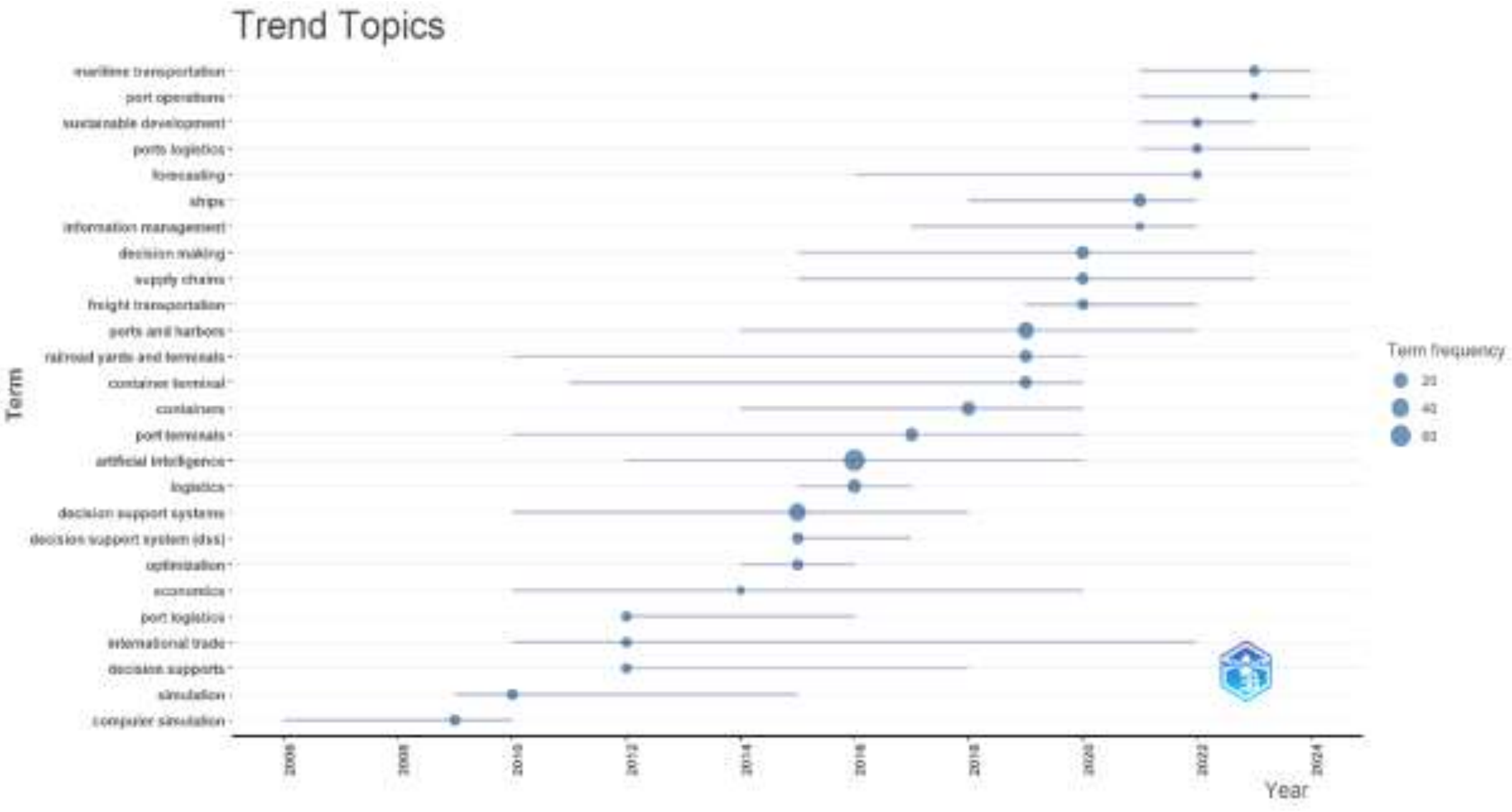

***Source: Authors***

### 4.12 Thematic Map

The conceptual landscape of the field is organized in Figure 12, offering a thematic classification based on development and centrality. The motor themes (upper right quadrant), such as “computer simulation”, “simulation”, “international trade”, and “port terminals”, are both highly developed and central to the field, indicating that they play a critical role in structuring the research area. The basic themes (lower right quadrant) include “artificial intelligence”, “decision support systems”, and “ports and harbors”, which are foundational and widely shared across studies but may still be maturing in terms of specialization and internal development.
The niche themes (upper left quadrant), like “machine learning”, “vessel”, and “5G mobile communication systems”, are highly developed but less connected to the broader research network, possibly reflecting highly technical subfields. On the other hand, emerging or declining themes (lower left quadrant), such as “environmental impact”, “passenger flow”, and “traffic congestion”, appear less developed and peripheral, raising questions about whether they represent new areas of interest or declining relevance.
Overall, the map reveals that current research in AI and port logistics is structured around operational optimization (motor themes) and foundational digital concepts (basic themes), while technical advances (like machine learning or IoT) and sustainability-related themes are still maturing or underexplored.

***Figure 12: Thematic Map***

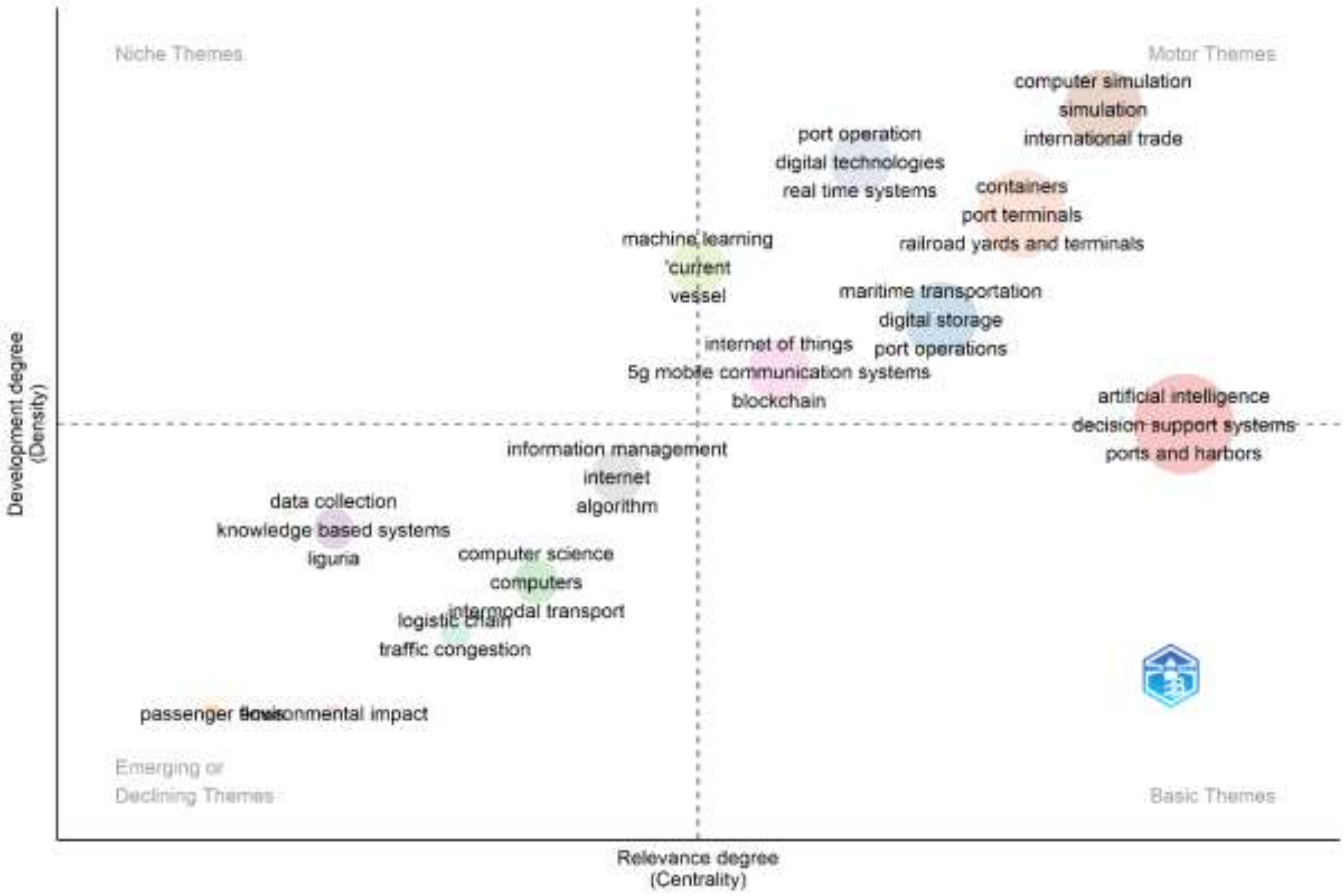

***Source: Authors***

### 4.13 Conceptual Structure Map

A deeper look into the intellectual structure is visualized in Figure 13, presenting keyword clusters identified through MCA). The map reveals three main thematic clusters. The left side of the map contains technical and operational terms which include machine learning and automatic identification systems and data mining and forecasting to show AI-based decision systems and automated port operations. The central section contains “logistics management” and “decision support systems” and “maritime logistics” and “supply chains” which represent the fundamental elements of this research domain. The established interdisciplinary knowledge base combines logistics with management and information systems through these specific terms. The right side of the map shows a cluster that includes “simulation” and “computer simulation” and “optimization” and “algorithms” which demonstrate the computational methods for optimizing port logistics systems. The group contains both general optimization methods and specific optimization approaches like “tabu search” and “genetic algorithms.” The research field maintains its structure through three interconnected elements which include AI methods and logistics management and simulation-based optimization and decision support that connects all clusters.

***Figure 13: Conceptual Structure Map***

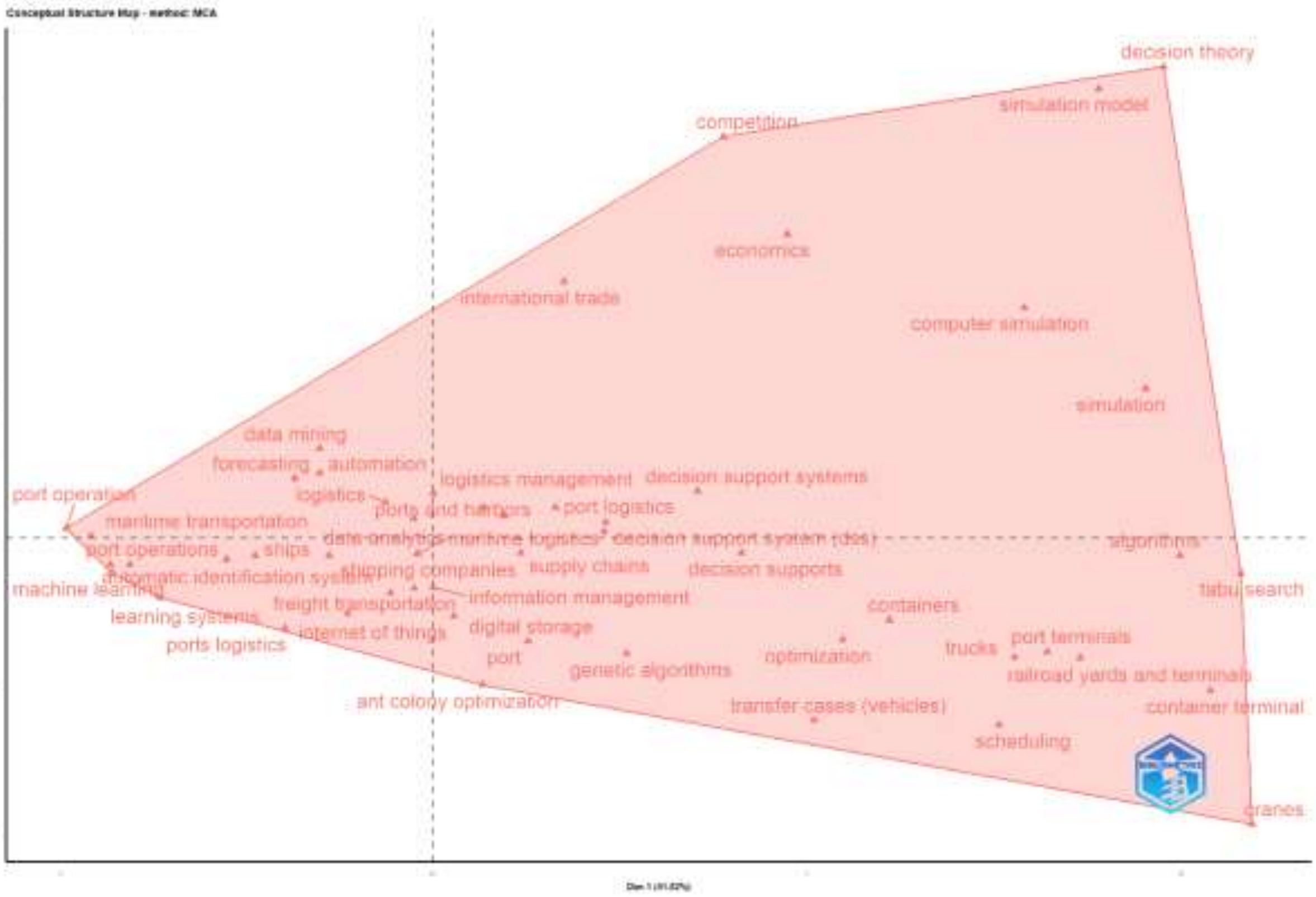

***Source: Authors***

## 5. Conclusion

This study has offered a detailed bibliometric analysis of the evolving academic landscape of artificial intelligence in maritime logistics and port operations. By analyzing 123 documents from the Scopus database using Bibliometrix, we mapped publication trends, key contributors, thematic clusters, and emerging research frontiers.

The findings underscore a rapid expansion of interest in AI applications, particularly in smart port development, environmental optimization, and digital twin technologies. Countries such as China, the USA, and South Korea are leading contributors, while core themes have shifted from basic automation to complex decision-support systems and sustainability-driven models.

Our research highlights not only the intellectual maturation of the field but also identifies critical gaps, especially the underrepresentation of empirical studies in low-income regions and limited integration of ethical and governance dimensions. The results serve as a valuable resource for researchers and practitioners seeking to understand the strategic trajectory of AI in maritime logistics.

Ultimately, this work contributes to advancing a structured research agenda that aligns with global goals for greener, smarter, and more resilient port systems in the era of digital transformation.

This study is not without limitations. First, it relies solely on the Scopus database, which, while comprehensive, may omit relevant studies indexed in other databases such as Web of Science or IEEE Xplore. Second, the exclusion of non-English publications may have led to an underrepresentation of contributions from non-Anglophone countries, especially those in Asia, South America, and Africa.

Additionally, bibliometric analysis captures patterns and structures in published research but does not assess the practical implementation or real-world impact of AI technologies in port environments. Therefore, future research could complement bibliometric methods with case studies or field surveys to provide a more holistic understanding.

Moreover, certain emerging areas such as AI ethics, cybersecurity in port systems, and AI-driven ESG (Environmental, Social, Governance) monitoring remain underexplored. These topics offer fertile ground for interdisciplinary research at the intersection of AI, maritime logistics, environmental science, and governance.

Future work should also prioritize longitudinal and comparative studies across ports with varying levels of digital maturity to better understand the conditions under which AI adoption succeeds or fails. Investigating policy and regulatory frameworks that facilitate or hinder the diffusion of AI innovations in the maritime sector would further enrich the academic discourse.

**References :**

(1). Benghalia, A., Ferdjallah, A., Oudani, M., & Boukachour, J. (2025). Machine learning and simulation for efficiency and sustainability in container terminals. *Sustainability, 17*, 2927.

(2). Bouman, E., et al. (2017). State-of-the-art technologies for greenhouse gas emissions reduction in shipping. Transportation Research Part D: Transport and Environment, 52, 408–421.

(3). Chang, Y., & Chen, C. (2020). Blockchain and AI convergence in maritime logistics: A conceptual framework. International Journal of Information Management, 52, 102059.

(4). Chang, Y.-C., & Chen, C.-F. (2021). Blockchain in maritime logistics: Challenges and opportunities. Maritime Policy & Management, 48(3), 332–347.

(5). Chen, H., Tang, Q., Jiang, Y., & Lin, Z. (2010). The role of international financial reporting standards in accounting quality: Evidence from the European Union. Journal of International Financial Management & Accounting, 21(3), 220–278.

(6). Chen, L., Yang, Y., & Tang, S. (2022). Artificial intelligence applications in smart ports: A review. Expert Systems with Applications, 204, 117594.

(7). Garcia, D., & Gonzalez, E. (2020). AI for port logistics optimization: The role of smart scheduling. Journal of Maritime Research, 17(2), 101–112.

(8). Gavalas, D., & Kasapakis, V. (2019). Real-time asset tracking in port logistics using AI and IoT. Sensors, 19(9), 2048.

(9). González-Cancelas, N., Palacios Calzada, J. P. C., Vaca-Cabrero, J., & Camarero-Orive, A. (2025). Optimizing sustainable port logistics in Spanish ports with emerging technologies. *Sustainability, 17*, 3392.

(10). Heilig, L., Lalla-Ruiz, E., & Voß, S. (2017). Digital transformation in maritime logistics: Smart ports in the making. Computers in Industry, 89, 16–30.

(11). Heilig, L., Schwarze, S., & Voss, S. (2017). An analysis of digital transformation in the history and future of modern ports. Electronic Markets, 27, 1–12.

(12). International Maritime Organization. (2020). Fourth IMO GHG Study 2020. International Maritime Organization.

(13). Ivanov, D. (2020). Viable supply chain model: Integrating agility, resilience and sustainability. International Journal of Production Research, 58(10), 2904–2915.

(14). Lee, J., Bagheri, B., & Kao, H. A. (2015). A cyber-physical systems architecture for Industry 4.0-based manufacturing systems. Manufacturing Letters, 3, 18–23.

(15). Lim, S., & Kim, H. (2023). Enhancing port efficiency with AI: A review of recent developments. Maritime Policy & Management, 50(3), 278–295.

(16). Liu, S., Zhang, Y., & Li, C. (2021). Predictive maintenance in port equipment using AI algorithms. Applied Sciences, 11(4), 1845.

(17). Luo, M., & Shin, S. H. (2022). The role of IoT and AI in port logistics: A comprehensive framework. Journal of Shipping and Trade, 7(1), 12.

(18). Lütjen, M., et al. (2021). Predictive analytics for container logistics disruption management. Journal of Supply Chain Management, 57(3), 42–57.

(19). Notteboom, T., & Rodrigue, J.-P. (2021). Port economics, management and policy. Routledge.

(20). Parola, F., & Satta, G. (2020). Digital technologies and logistics: A research agenda. Transportation Research Part E: Logistics and Transportation Review, 142, 102067.

(21). Psaraftis, H. N., & Kontovas, C. A. (2020). Speed optimization and emissions in ECAs: A trade-off analysis. Maritime Economics & Logistics, 22(1), 73–95.

(22). Psaraftis, H., & Zis, T. (2018). The economics of speed reduction in maritime transport. Transportation Research Part D: Transport and Environment, 62, 298–307.

(23). Shukla, N., & Jharkharia, S. (2013). Applicability of artificial intelligence in green supply chain management: A literature review. International Journal of Logistics Systems and Management, 14(4), 393–410.

(24). Tao, F., et al. (2019). Digital twins and cyber–physical systems in smart manufacturing. Engineering, 5(4), 653–661.

(25). Wu, H., Li, W., & Shi, W. (2023). Digital twin-enabled smart maritime logistics 5.0: Architecture, applications and challenges. Ocean Engineering, 270, 113361.

(26). Zhang, L., Zhang, H., & Chen, J. (2022). Artificial intelligence in underwater digital twins. Journal of Marine Science and Engineering, 10(3), 322.

(27). Zhang, R., & Lee, C. (2022). AI-based early warning systems for port logistics. Maritime Transport Review, 9(1), 45–62.

(28). Zhu, S., & Wang, W. (2024). A bibliometric analysis of AI applications in global logistics. Journal of Business Research, 160, 113752.